\documentclass[ 
    aps,
    pra,
    twocolumn,
    letterpaper, 
    10pt,
    superscriptaddress, 
    showpacs,
    showkeys,
    notitlepage,
    amsmath, 
    amssymb, 
    floatfix 
]{revtex4-1}

\usepackage{graphicx}
\usepackage{dcolumn}
\usepackage{bm}
\usepackage{color}
\usepackage{amssymb}
\usepackage{amsmath}
\usepackage{amsbsy}
\usepackage{hyperref}
\usepackage{bbold}
\usepackage{accents}

\newcommand{\harpvecsign}{\scriptscriptstyle\text{\tiny$\leftrightarrow$}}
\newcommand{\harpoonvec}[2]{%
  \ifx\displaystyle#1\doalign{$\harpvecsign$}{#1#2}\fi
  \ifx\textstyle#1\doalign{$\harpvecsign$}{#1#2}\fi
  \ifx\scriptstyle#1\doalign{\scalebox{.6}[.9]{$\harpvecsign$}}{#1#2}\fi
  \ifx\scriptscriptstyle#1\doalign{\scalebox{.5}[.8]{$\harpvecsign$}}{#1#2}\fi
}
\newcommand{\doalign}[2]{%
 {\vbox{\offinterlineskip\ialign{\hfil##\hfil\cr#1\cr$#2$\cr}}}%
}

\newcommand{\bel}{\begin{equation}}
\newcommand{\eel}{\end{equation}}

\newcommand{\skyp}[1]{}

\newcommand{\fr}{\frac}

\newcommand{\vare}{\varepsilon}

\newcommand{\ee}{\end{equation}}
\newcommand{\be}{\begin{equation}}
\newcommand{\mbf}{\mathbf}

\newcommand{\bal}{\begin{eqnarray} }
\newcommand{\eal}{\end{eqnarray}}
\newcommand{\ba}{\begin{eqnarray*}}
\newcommand{\ea}{\end{eqnarray*}}

\newcommand{\reffig}[1]{Fig.~\ref{#1}}

\newcommand{\ket}[1]{| #1 \rangle}

\newcommand{\bp}{{\mathbf p}}
\newcommand{\br}{{\mathbf r}}

\newcommand{\bR}{{\mathbf R}}

\newcommand{\bk}{{\mathbf k}}

\newcommand{\bK}{{\mathbf K}}

\newcommand{\refeq}[1]{Eq.~\eqref{#1}}

\begin{document} 

\title{Theory of Dipole Radiation Near a Dirac Photonic Crystal}

\author{J. Perczel}
\email[email: ]{jperczel@mit.edu}
\affiliation{Physics Department, Massachusetts Institute of Technology, Cambridge, MA 02139, USA}
\affiliation{Physics Department, Harvard University, Cambridge,
MA 02138, USA}


\author{M. D. Lukin}
\affiliation{Physics Department, Harvard University, Cambridge,
MA 02138, USA}


\date{\today}

\bigskip
\bigskip
\bigskip
\begin{abstract} 

We develop an analytic formalism to describe dipole radiation near the Dirac cone of a two-dimensional photonic crystal slab. In contrast to earlier work, we account for all polarization effects and derive a closed-form expression for the dyadic Green's function of the geometry. Using this analytic Green's function, we demonstrate that the dipolar interaction mediated by the slab exhibits winding phases, which are key ingredients for engineering topological systems for quantum emitters, as discussed in detail in {\color{blue} J. Perczel et al., (2018), arXiv:1810.12299}. 
As an example, we study the coherent atomic interactions mediated by the Dirac cone, which were recently shown to be unusually long-range with no exponential attenuation. These results pave the way for further, rigorous analysis of emitters interacting in photonic crystals via photonic Dirac cones.


\end{abstract}

\maketitle


\section{Introduction}

Since the experimental discovery of single-atom graphene sheets \cite{Novoselov2004}, the Dirac-like energy spectrum of graphene has been studied extensively \cite{Neto2009}. This has inspired research into finding photonic analogues of graphene with Dirac-like dispersion for photons, which have been shown to give rise to a number of remarkable phenomena, including topological waveguides at microwave and optical frequencies \cite{Haldane2008,Raghu2008,Barik2016,Barik2018}, pseudo-diffusive transport of light \cite{Sepkhanov2007,Zandbergen2010}, Klein tunneling \cite{Bahat-Treidel2010} and photonic Zwitterbewegung \cite{Zhang2008}. Photonic Dirac cones have also been studied in the context of quantum optics and were found to give rise to single-mode behavior over large areas \cite{Bravo-Abad2012} and two-dimensional localization without a band gap  \cite{Xie2014}. More recently, quantum emitters coupling to photonic Dirac cones were found to exhibit exotic quantum dynamics and purely long-range coherent interactions without exponential attenuation \cite{Gonzalez-Tudela2018}.

A key challenge for making such devices useful for quantum optical applications is to find realistic structures, in which the Dirac cone is energetically detuned from all other bands \cite{Barik2016} and which support quasi-two-dimensional propagation of light \cite{Bravo-Abad2012}. An additional challenge is the inherent complexity of the electromagnetic spectrum of such devices, which usually makes full-scale numerical analysis onerous \cite{Bravo-Abad2012,Barik2016,Gonzalez-Tudela2018}. Consequently, previous studies of the photonic Dirac cone relied on simplified analytic models that neglected key aspects of the electromagnetic properties of the devices, such as the polarization structure of the modes. Therefore, it is of significant interest to investigate whether there exist realistic quasi-two-dimensional structures with photonic Dirac cones, whose electromagnetic spectrum -- including polarization effects -- can be understood analytically.   

\begin{figure}[h!]
\centering
\includegraphics[width=0.475 \textwidth]{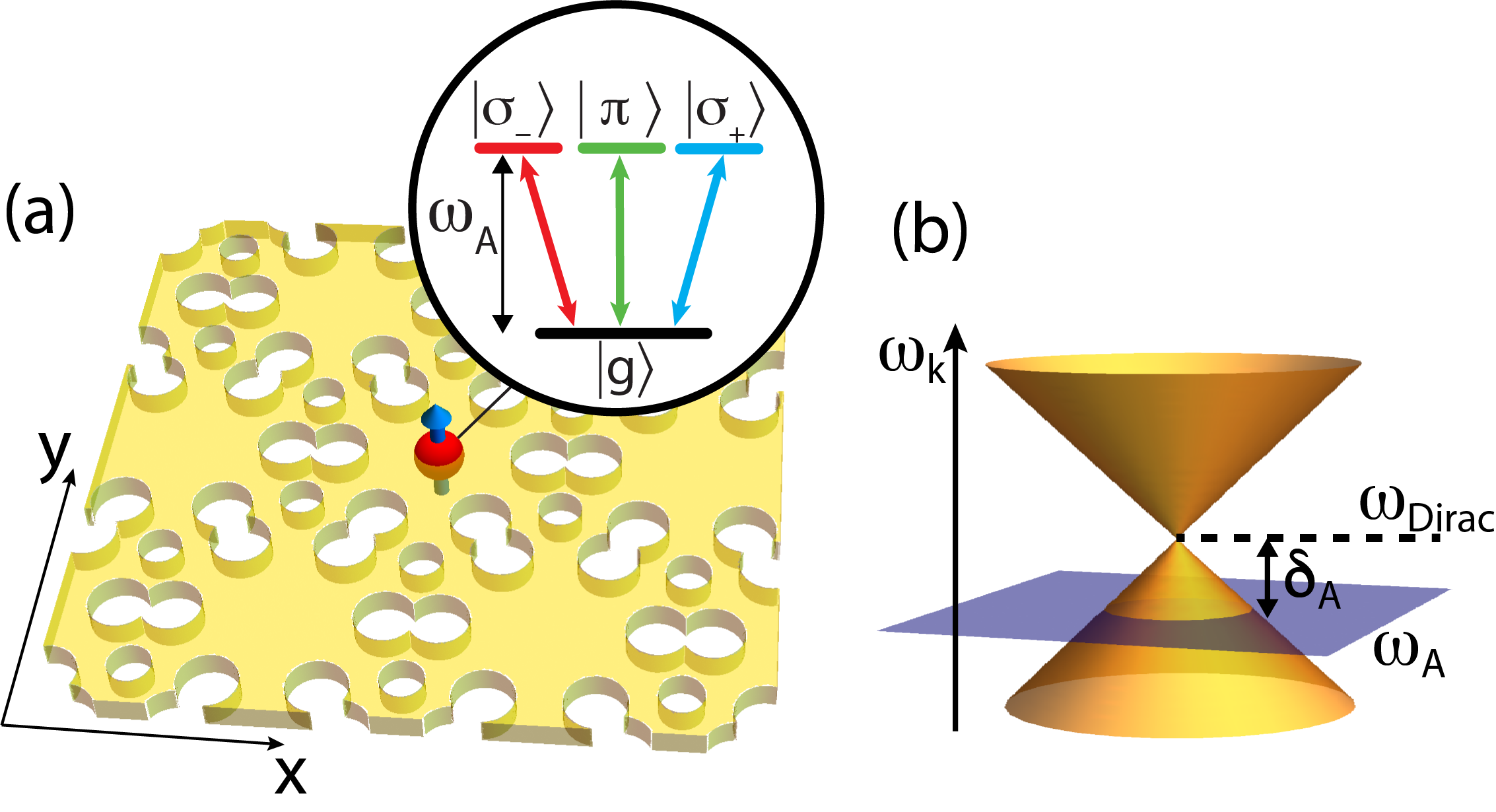}
\caption{
\label{fig:schematics}
(a) Quantum emitter embedded in a quasi-two-dimensional photonic crystal slab with a cavity-like unit cell. The emitter is assumed to have three degenerate transitions of energy $\omega_A$ from the ground $\ket{g}$ to the excited states $\ket{\sigma_\pm}$ and $\ket{\pi}$. (b) The photonic bands of the slab feature a Dirac-like dispersion cone. The emitter transition frequency $\omega_A$ is detuned by $\delta_A$ from the vertex of the Dirac cone $\omega_\text{Dirac}$. } 
\end{figure} 

In this work, we develop an analytic formalism to describe dipole radiation near a realistic photonic crystal slab with a Dirac cone (\reffig{fig:schematics}(a)). The Dirac cone is energetically detuned from all the other bands, allowing us to tune the transition frequency of quantum emitters close to the Dirac point (i.e. the Dirac cone vertex) without interacting with any other bands, as shown in \reffig{fig:schematics}(b). Consequently, all dipolar interactions are mediated by the modes comprising the photonic Dirac cone, simplifying the form of the interaction. Crucially, we find that the polarization structure of the modes of the Dirac cone can be captured by an analytic model, which allows us to incorporate polarization effects into our quantum optical treatment. Using our analytic model, we obtain a Green's function in a closed form, which captures all cooperative atomic effects relevant for the quantum dynamics of two or more emitters. In addition, we also find that the dipolar coupling between emitters exhibits winding phases, which are the key ingredients for engineering topological atomic arrays \cite{Perczeletal2018}. Finally, as an application of our formalism, we study the coherent atomic interactions mediated by the photonic Dirac cone, which were recently shown to be unusually long-range with no exponential attenuation \cite{Gonzalez-Tudela2018}.



This paper is organized as follows. In Sec.~\ref{PcStructure} we describe how to engineer a two-dimensional photonic crystal slab with a Dirac-like dispersion that is energetically fully separated from the rest of the bands. In Sec.~\ref{DiracConeModel} we develop an analytic model for the photonic modes that constitute the Dirac cone. In Sec.~\ref{GreensFunction} we calculate the Green's function that describes the dipolar coupling between emitters. In Sec.~\ref{InteractionsAnalysis} we analyze the interactions mediated by the Dirac cone. Key results and conclusions are presented in Sec.~\ref{Conclusion}.

%
%

\section{Photonic crystal slab with a Dirac dispersion}\label{PcStructure}

In this section, we discuss how to engineer a quasi-two-dimensional photonic crystal slab that features bands with Dirac-like conical dispersion, such that the Dirac cone is energetically detuned from all other bands.

\begin{figure}[h!]
\centering
\includegraphics[width=0.475 \textwidth]{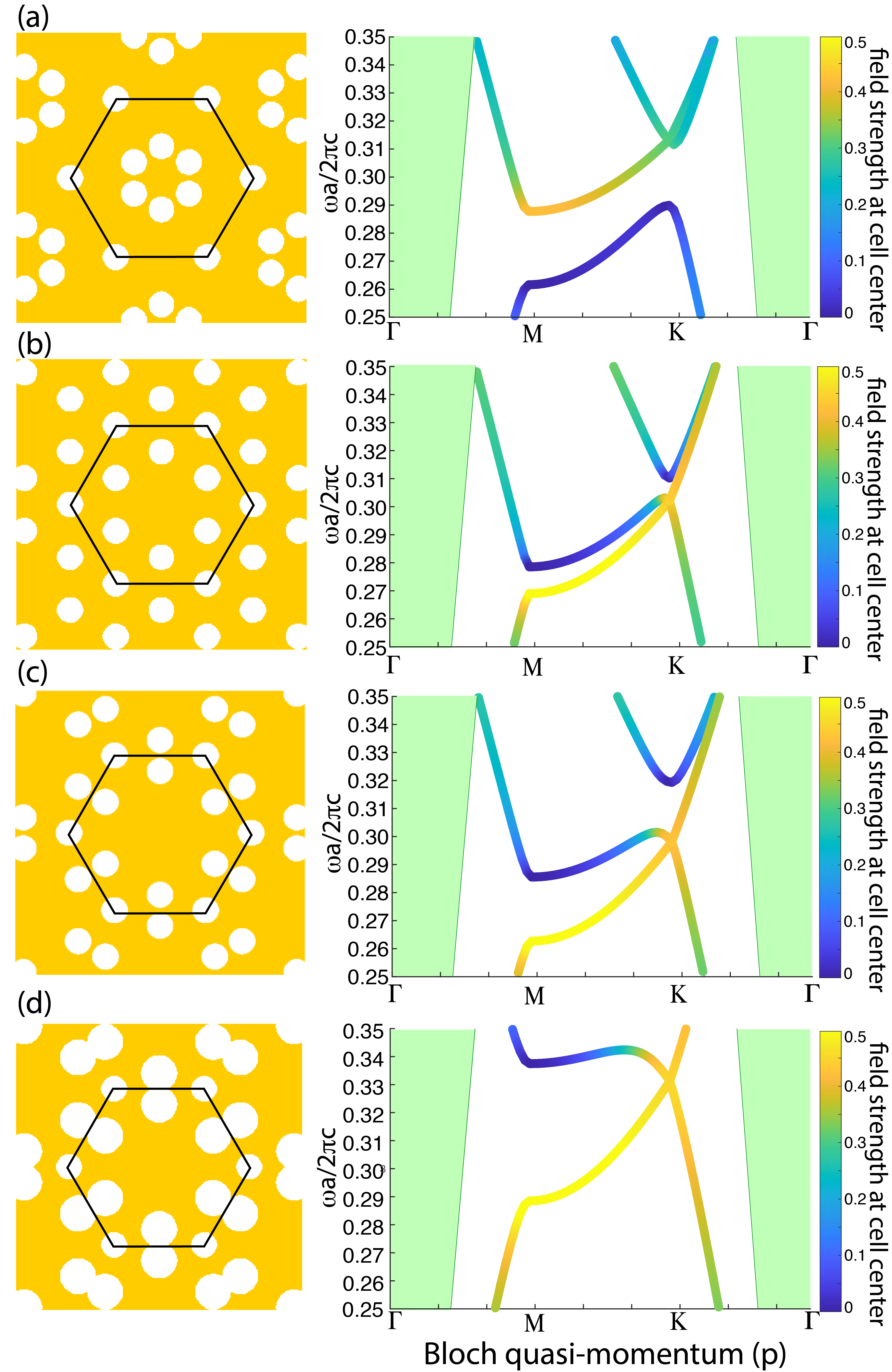}
\caption{
\label{fig:phasetransition}
Photonic crystal slab unit cell and the corresponding TE-like band structure for 4 different geometries. The thickness of the slab is $h=0.5a$, the radius of the holes is $r_s=0.0833a$ and the dielectric permittivity is $\vare_\text{GaP}=10.5625$. (a) When 6 holes are placed $l=0.2a$ from the center, the modes in the upper band have a higher field intensity than in the lower band. Color code shows the electric field strength $|\mbf E_\mbf{p}(\mbf 0)|^2a^3$ at the center of the cell for each mode. (b) For $l=0.333a$ the modes in the lower band have a higher field intensity than in the upper band. (c) Pushing the holes to the edge of the cell ($l=0.4a$) increases the energy gap between the bands at the $\mbf M$ point. (d) Increasing the radius of the 6 holes to $r_l=0.116a$ makes the Dirac cone energetically separated from all other bands. } 
\end{figure}

The key idea is illustrated in \reffig{fig:phasetransition}. We consider a gallium phosphate (GaP) photonic crystal of finite thickness $h$ with air holes. The hole pattern of the slab has an underlying triangular lattice symmetry with lattice constant $a$ and lattice vectors 
\be\label{LatticeVectors}
{\mbf a_\pm = a(\pm \sqrt{3}\hat x+\hat y)/2}.
\ee
The hexagonal unit cell of the triangular lattice is comprised of 6 holes with radius $r_l$ near the center of the cell and an additional set of holes with radius $r_s$ on the boundaries, which fall only partially inside the unit cell. \reffig{fig:phasetransition}(a)-(c) shows the holes for three different values of $l$, where $l$ is the radial distance of the 6 holes from the center of the cell. We use the plane-wave expansion method to numerically solve Maxwell's equations \cite{Johnson2001} and obtain the lowest-lying transverse electric (TE-like) bands for the unit cells shown. Here, we only consider TE-like modes, since in what follows, we will assume that the quantum emitters sit at the center of the unit cell at $\br = \mbf 0$, where tranverse magnetic (TM-like) modes have zero field strength due to the inversion symmetry of the structure \cite{Joannopoulos2008}. This symmetry ensures that the $z$-component of the TE-like modes is also zero at $\br = \mbf 0$. The color code for the bands shows the normalized electric field strength $|\mbf E_\mbf{p}(\mbf 0)|^2a^3$ of each mode, sampled at the center of the cell ($\br = \mbf 0$).  

From the colors of the bands in \reffig{fig:phasetransition}(a), we see that when the 6 holes are close to each other, the modes in the lowest band have a low field concentration in the dielectric at the cell center, whereas the upper band has a high field concentration. As the holes are pushed radially outward, the two bands approach each other, and eventually cross, and thus the lower band has higher field concentration (\reffig{fig:phasetransition}(b)). When the holes are pushed radially further out, the two lowest bands energetically separate at the $\mbf M$ point, while maintaining their degeneracy at the $\mbf K$ point, forming a Dirac cone as shown in \reffig{fig:phasetransition}(c). Once the holes reach the edge of the irreducible Brillouin zone, the band separation at the $\mbf M$ point can still be increased by expanding the radii of the 6 holes as shown in \reffig{fig:phasetransition}(d). For sufficiently large radii, the modes of the air band are no longer degenerate with the modes of the Dirac cone. Thus, the Dirac cone becomes energetically well-separated from all other bands, making it possible to tune emitters close to the Dirac point without coupling to any modes other than those constituting the Dirac cone. Note that similar techniques have been used previously to control the gap at the photonic Dirac cone for topological waveguiding with a single quantum emitter \cite{Barik2016,Barik2018}. 

In what follows, we will assume that the transition frequency of the emitters $\omega_A$ is tuned sufficiently close to the frequency of the Dirac point $\omega_\text{Dirac}$ (i.e. ${\omega_\text{Dirac}-\omega_A\ll \omega_A}$) that coupling to all modes away from the Dirac cone is small and can be neglected.

\section{Analytic model for the Dirac cone}\label{DiracConeModel}

In this section, we develop an analytic approximation to the electromagnetic modes that constitute the photonic Dirac cone. This will enable us to derive a closed-form expression for the dyadic Green's function of the geometry (see Sec.~\ref{GreensFunction}), which describes the atomic properties inside the photonic crystal.  

We start by parameterizing the linear dispersion of the Dirac cone as a function of the quasi-momentum $\bp$ as
\be\label{Dispersion}
\omega_{\bp}^{(\mbf{K},\pm)} = \omega_\text{Dirac}\pm v_\text{s}|\bp-\bp_\mbf{K}|,
\ee
where $(+)$ and $(-)$ correspond to the top and bottom bands respectively, $\omega_\text{Dirac}$ is the energy associated with the vertex of the cone, $\bp_\mbf{K}=2\pi/a(1/\sqrt{3},1/3)$ is the momentum vector of the Dirac cone inside the irreducible Brillouin zone and $v_\text{s}$ gives the group velocity of the guided modes in units of $c$. Mathematically, $v_\text{s}$ corresponds to the slope of the Dirac cone. Both $\omega_\text{Dirac}$ and $v_\text{s}$ can be obtained from numerical calculations. 

\begin{figure}[h!]
\centering
\includegraphics[width=0.475 \textwidth]{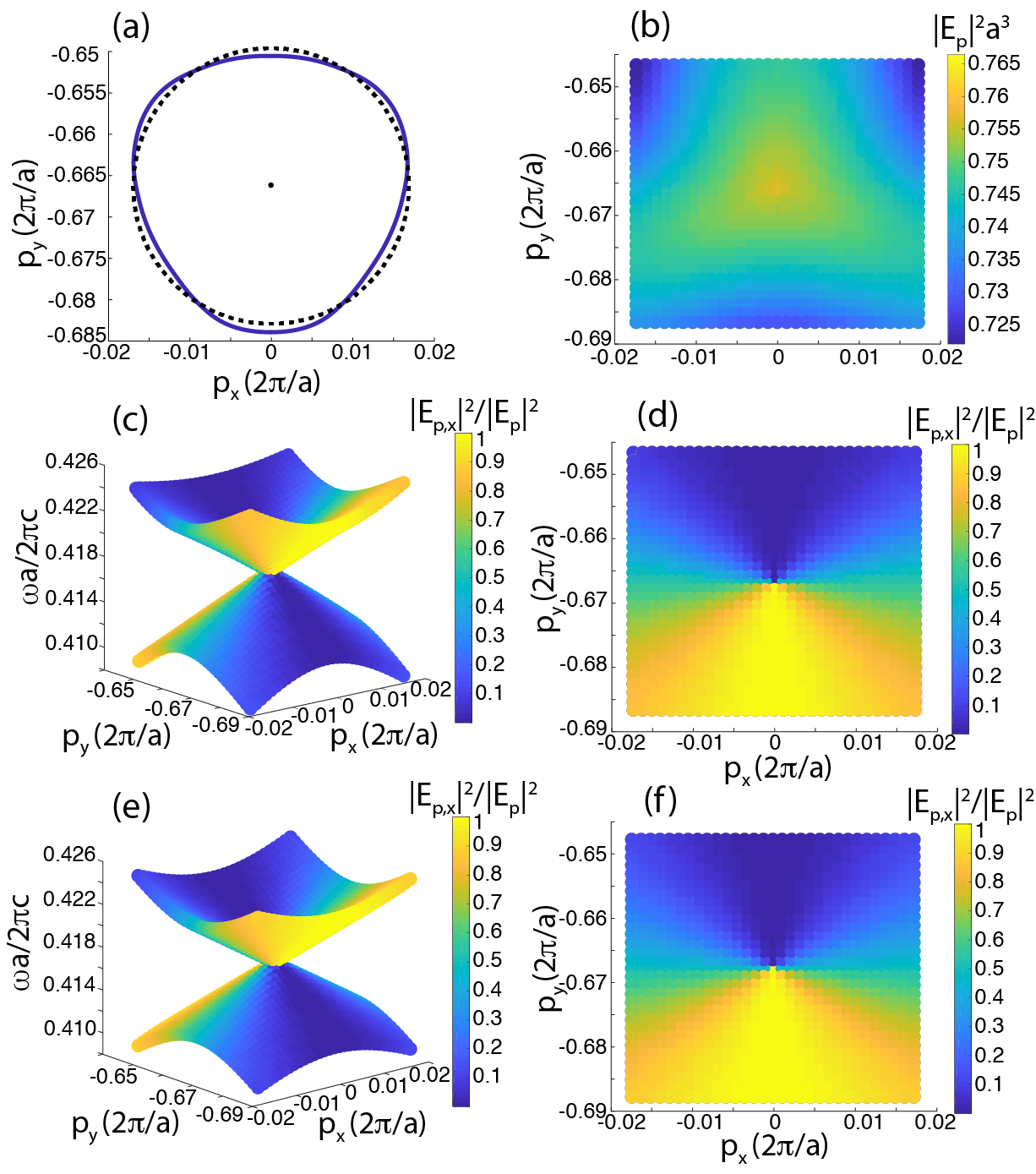}
\caption{
\label{fig:analytic}
(a) Cross section of the Dirac cone for $\omega a/2\pi c=0.422$. Numerical results are shown in solid blue, analytic prediction is shown in dashed black. (b) Field intensity $|\mbf E_\bp(\mbf 0)|^2 a^3$ in the top band near the Dirac cone. All values fall in a small interval of $[0.725,0.765]$. (c) Numerically obtained results for the Dirac bands. Color code shows the $x$-component of the field $|E_{\bp,x}(\mbf 0)|^2/|\mbf E_{\bp}(\mbf 0)|^2$. (d) Same as (c), showing the top band from above. (e) \& (f) Analytic approximation to the band dispersion and the polarization of the modes. Numerical results were obtained for a GaP structure with geometric parameters $h=0.25a$, $r_s=0.0833a$, $r_l = 0.15a$ and $l=0.4a$, for which we obtain the values $\omega_\text{Dirac}a/2\pi c=0.4172$ and $v_\text{s}=0.3c$.} 
\end{figure}

While the dispersion is linear in the immediate vicinity of the Dirac cone, the slope $v_\text{s}$ is found to vary slightly along different directions. \reffig{fig:analytic}(a) shows the cross section of the cone taken above the Dirac point. The black dashed line shows the results predicted by \refeq{Dispersion}, while the solid blue line gives the numerically obtained results showing slight deviations from perfect circularity due to the underlying three-fold symmetry of the lattice. We average the numerical values of $v_\text{s}$ along the different directions to obtain an effective value for $v_\text{s}$, which we can use in \refeq{Dispersion}.   

Next, we explore the polarization structure of the electromagnetic modes at the center of the unit cell. \reffig{fig:analytic}(b) shows the overall field intensity $|\mbf E_\bp(\mbf 0)|^2 a^3$ of each mode in the top band at $\br=\mbf 0$. The intensity varies only weakly as a function of momentum and falls within a small interval. Similar results are obtained in the lower band. Therefore, we average the field intensity of all the modes to obtain an effective constant value $|E_0|^2a^3$ that we use to parameterize the field intensity of all modes in the vicinity of the Dirac cone.  

To understand the contributions of the different polarization components to the overall intensity, in \reffig{fig:analytic}(c) we color each of the modes of the Dirac cone according to the normalized intensity of the $x$-component of the field $|E_{\bp,x}(\mbf 0)|^2/|\mbf E_{\bp}(\mbf 0)|^2$. \reffig{fig:analytic}(d) shows the colored modes in the top band from above. The simple functional dependence of the resulting intensity pattern can be approximated by $\sin^2(\Phi_{\mbf K}/2-\pi/4)$ for the top band and $\sin^2(\Phi_{\mbf K}/2+\pi/4)$ for the bottom band, where the polarization is expressed as a function of the angle $\Phi_{\mbf K}$ measured from the $p_x$ axis. Figs.~\ref{fig:analytic}(e) and (f) show the corresponding results obtained from this simple analytic model and using \refeq{Dispersion}. Good agreement is obtained between the numerical and analytic results. Similarly, for the $y$-component of the electric field $|E_{\bp,y}(\mbf 0)|^2$ we find that the change in polarization as a function of $\Phi_{\mbf K}$ is captured by $\sin^2(\Phi_{\mbf K}/2+\pi/4)$ for the top band and $\sin^2(\Phi_{\mbf K}/2-\pi/4)$ for the bottom band.

In general, we can model the electric field of the Dirac cone modes at $\br=\mbf 0$ as 
\bal\label{PolarizationStructureK}
\mbf E^{(\mbf K,\pm)}_{\bp} = E_0\left[ \sin\left(\fr{\Phi_{\mbf K}}{2}\mp \fr{\pi}{4}\right)\hat x \pm \sin\left(\fr{\Phi_{\mbf K}}{2}\pm\fr{\pi}{4}\right)\hat y \right]\!,\quad
\eal  
where
\bal\label{PhiK}
\Phi_{\mbf K} (p_x,p_y)= \arctan\left(\fr{p_y-p_{\mbf K,y}}{p_x-p_{\mbf K,x}}\right).
\eal
Note that there are two inequivalent Dirac cones inside the irreducible Brillouin zone, and the dispersion at the second Dirac cone can be expressed as
\be\label{Dispersion2}
\omega_{\bp}^{(\mbf K',\pm)} = \omega_\text{Dirac}\pm v_\text{s}|\bp-\bp_\mbf{K'}|,
\ee
where $\bp_{K'}=2\pi/a(1/\sqrt{3},-1/3)$ is the momentum vector associated with the second Dirac cone. As in the case of the other Dirac cone, we find that the electric field of the modes of the second Dirac cone at $\br = \mbf 0$ can be expressed as
\bal\label{PolarizationStructureKprime}
\mbf E^{(\mbf K',\pm)}_{\bp} = E_0\left[ \sin\left(\fr{\Phi_{\mbf K'}}{2}\pm \fr{\pi}{4}\right)\hat x \mp \sin\left(\fr{\Phi_{\mbf K'}}{2}\mp\fr{\pi}{4}\right)\hat y \right]\!,\quad\;
\eal  
where
\bal\label{FieldKprime}
\Phi_{\mbf K'} (p_x,p_y)= \arctan\left(\fr{p_y-p_{\mbf K',y}}{p_x-p_{\mbf K',x}}\right).
\eal

\begin{figure}[h!]
\centering
\includegraphics[width=0.475 \textwidth]{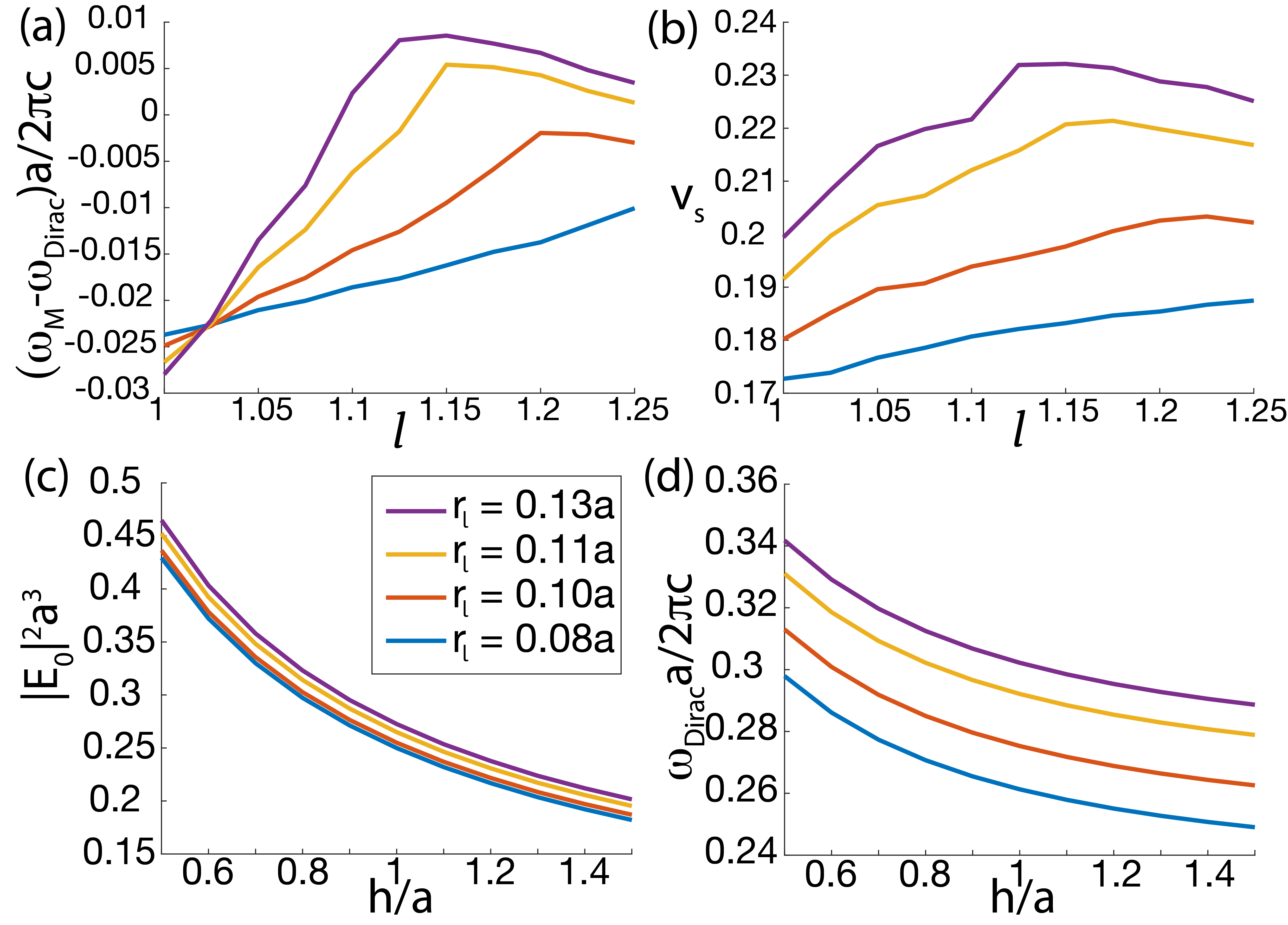}
\caption{
\label{fig:greens}
(a) The detuning $\omega_{\mbf M}-\omega_\text{Dirac}$ as a function of $r_l$ and $l$, the radius and radial displacement of the central holes. Slab height is $h=1.13a$. (b) The slope constant $v_\text{s}$ as a function of $r_l$ and $l$ for $h=1.13a$. (c) The electric field strength $|E_0|^2$ as a function of $r_l$ and the slab thickness $h$ for $l=0.4a$. (d) The frequency of the Dirac point $\omega_\text{Dirac}$ as a function of $r_l$ and $h$ for $l=0.4a$. The inset of (c) provides the color-coded legend for the curves with different hole radii. } 
\end{figure}

The simple analytic model descibed by Eqs.~\eqref{Dispersion}-\eqref{FieldKprime} shows that the electromagnetic environment of an emitter tuned close to the vertex of the Dirac cone is completely described by the parameters $\omega_\text{Dirac}$, $v_\text{s}$ and $E_0$, as long as the second band near the $\mbf M$ symmetry point is energetically separated from the Dirac point (see \reffig{fig:phasetransition}), i.e. $\omega_{\mbf M}>\omega_\text{Dirac} $. In \reffig{fig:greens} we plot these four numerical parameters as a function of the geometric parameters $r_l$, $l$ and $h$. More specifically, we plot $\omega_{\mbf M}-\omega_\text{Dirac}$ and $v_\text{s}$ only as function of $r_l$ and $l$, since we found that these quantities are largely insensitive to $h$. For the same reason, we plot $|E_0|^2$ and $\omega_\text{Dirac}$ only as a function of $r_l$ and $h$, but not $l$.

 \reffig{fig:greens}(a) shows that $\omega_{\mbf M}-\omega_\text{Dirac}$ initially increases with $l$, before decreasing in magnitude.  Intuitively, once the holes reach the edges of the unit cell increasing $l$ does not help any more. Instead, pushing the holes further out merges them together, reducing the volume of air holes. The figure also shows that $\omega_{\mbf M}-\omega_\text{Dirac}$ can be increased by increasing the hole radius.  Similarly, \reffig{fig:greens}(b) shows that the slope of the Dirac cone $v_\text{s}$ increases both with $l$ (until the holes reach the edge of the unit cell) and the hole radius.
 
\reffig{fig:greens}(c) shows that the electric filed intensity is largely insensitive to the hole radius, but varies inversely with $h$. This latter result makes intuitive sense -- increasing the thickness of the slab expands the mode volume and, therefore, decreases the field concentration at the location of the emitters. Finally,  \reffig{fig:greens}(d) shows that $\omega_\text{Dirac}$ is inversely proportional to $h$ and increases with hole radius. These results are simple consequences of the fact that increasing the relative volume of the dielectric inside the unit cell increases the effective refractive index of the medium and thus rescales all relevant length scales of the system, including the effective wavelength.

\section{Calculation of the Green's function}\label{GreensFunction}

In quantum optics, single atom decay rates and cooperative effects such as the dipole-dipole interaction and cooperative decay of atoms can be fully determined from the dyadic Green's function of the problem when the Born-Markov approximation holds \cite{Dzsotjan2010,Dzsotjan2011,Perczel2018,Gonzalez-Tudela2015,Gonzalez-Tudela2018}. In particular, the overall spontaneous decay rate of an atom can be expressed as
\bal\label{spontaneousdecay}
\Gamma = \fr{2d^2\omega_A^2}{\hbar \vare_0c^2}\text{Im}\{G_{\alpha\alpha}(\mbf 0)\},
\eal
where $d$ is the transition dipole moment of the atom, $\vare_0$ is the electric permittivity in vacuum, $\alpha = x,y,z$ and the Green's function is evaluated at the location of the emitter ($\br = \mbf 0$). Similarly, the dipole-dipole interaction and cooperative decay rate between two atoms at a distance $\br$ apart are given by 
\bal\label{coop}
\delta\omega_\text{coop}(\br) = \fr{d^2\omega_A^2}{\hbar \vare_0c^2}\text{Re}\{G_{\alpha\beta}(\br)\},
\eal
and
\bal\label{dd}
\Gamma_\text{coop} (\br)= \fr{2d^2\omega_A^2}{\hbar \vare_0c^2}\text{Im}\{G_{\alpha\beta}(\br)\},
\eal
respectively, where the first atom is assumed to be in its $\alpha$-polarized excited state and the second atom is in its $\beta$-polarized state ($\alpha,\beta = x,y,z$). Note that the single-excitation atomic dynamics can be fully determined in terms of $\Gamma$, $\delta\omega_\text{coop}$ and $\Gamma_\text{coop}$, when the Born-Markov approximation is valid (see Ref.~\cite{Gonzalez-Tudela2018} for discussion of the case when these approximations break down). In this section we will derive an analytic approximation for the dyadic Green's function. 

Before proceeding, we note that since the $z$-component of the TE-like modes of the slab is zero at the location of the emitters (see Section ~\ref{PcStructure}), the $\ket{\pi}$ states of the emitter do not couple to the the guided modes and thus 
\bal
G_{zz}=G_{xz}=G_{zx}=G_{yz}=G_{zy}=0.
\eal
Therefore, we will focus our attention on the $x$- and $y$-polarized states of the emitters and derive the Green's function for these states.  


We start by expessing the momentum-space Green's function as a summation over the electric field eigenfunctions of the geometry \cite{Perczel2018} 
\be\label{GreensMomentum}
g_{\alpha\beta}(\bp) =\mathcal{A}c^2 \sum\limits_{n}\fr{E_{\bp,\alpha}^{(n)*}(\br_1)\,E_{\bp,\beta}^{(n)}(\br_2)}{\omega_A^2-\big(\omega_\bp^{(n)}\big)^2},
\ee
where $\mathcal A=\sqrt{3}/2a^2$ is the area of the hexagonal unit cell, $E^{(n)}_{\bp,\alpha}(\br)$ denotes the $\alpha$ component ($\alpha=x,y$) of the electric field of the photonic crystal mode in the $n^\text{th}$ band for quasi-momentum $\bp$ at location $\br$, $\omega_\bp^{(n)}$ is the corresponding frequency of the mode and the summation runs over all bands. In the expression above $\br_1$ is the position of the emitter and $\br_2$ is the position where the Green's function is evaluated (e.g. the location of the second atom). 

In order to find the Green's function in real space, it is necessary to integrate \refeq{GreensMomentum} across the entire irreducible Brillouin zone
\bal\label{integral1}
G_{\alpha\beta}(\br_2) = \int_\text{BZ} \fr{d^2\bp}{(2\pi)^2}g_{\alpha\beta}(\bp).
\eal
To perform this integral, we need to make a number of approximations.

First, note that given to the periodicity of the photonic crystal, the electric field eigenmodes can be expressed in the following Bloch form 
\bal\label{normalizedU}
E^{(n)}_{\bp,\alpha}(\br) = \fr{u^{(n)}_{\bp,\alpha}(\br)}{\sqrt{a^3}}e^{\text{i}\bp\cdot \br},
\eal
where ${\mbf u^{(n)}_\bp(\br+\bR)=\mbf u^{(n)}_\bp(\br)}$ is a periodic and dimensionless function and $\bR = n\, \mbf a_+ + m\, \mbf a_-$ is any valid lattice vector of the triangular lattice, where $n,m \in \mathbb{Z}$ and $\mbf a_\pm$ are defined in \refeq{LatticeVectors}. The eigenmodes are normalized such that  
\bal
\int_\mathcal{V}d^3\br\; \vare(\br)\,  \mbf{E}_{\bp}^{(n)}(\br)\cdot \mbf{E}_{\bp'}^{(n)*}(\br) = \delta_{\bp\bp'},
\eal
where $\vare(\br+\bR)=\vare(\br)$ describes the periodic dielectric permittivity of the photonic crystal in real space and the integral is performed over the quantization volume $\mathcal{V}$ \cite{Glauber1991,Perczel2018}. We also assume that $\br_2=\br_1+\bR$, which allows us to write 
\bal
E_{\bp,\alpha}^{(n)*}(\br_1)\,E_{\bp,\beta}^{(n)}(\br_2) =\fr{u_{\bp,\alpha}^{(n)*}(\br_1)\,u_{\bp,\beta}^{(n)}(\br_1)}{a^3} e^{i\bp\cdot \br_{12}},\quad
\eal
where we have defined $\br_{12}=\br_1-\br_2$. In addition, recall our assumption that $\omega_\text{Dirac}-\omega_A\ll \omega_A$, which ensures that only the bands of the two inequivalent Dirac cones at $\mbf K$ and $\mbf K'$ contribute to the integral. Therefore, we can make the approximation 
\be
\omega_A^2-(\omega_{\bp }^{(n)})^2 \approx 2\omega_A(\omega_A-\omega_{\bp}^{(n)}).
\ee
Finally, we assume that $\br_1$ (and hence $\br_2$) is positioned at the center of the unit cell, where Eqs.~\eqref{PolarizationStructureK}-\eqref{FieldKprime} are valid.

With these assumptions, the integral \refeq{integral1} can be evaluated in a closed form (see Appendix) and we obtain
\bal\label{GreensCartesian1}
G_{xx}(r,\phi)&=&  \sin\phi\, P^{-}(\br)  H_1^{(2)}( r/\xi)\nonumber\\
&&-iP^{+}(\br) H_0^{(2)}(r/\xi),\quad\;\,\,\\
G_{yy} (r,\phi)&=& -\sin\phi\, P^{-}(\br)  H_1^{(2)}( r/\xi)\nonumber\\
&&-iP^{+}(\br) H_0^{(2)}(r/\xi),\quad\;\,\,\\
G_{xy} (r,\phi)&=&  \cos\phi \, P^{-}(\br)H_1^{(2)}( r/\xi),\quad\;\\
G_{yx} (r,\phi)&=& G_{xy} (r,\phi),\quad\;\label{GreensCartesian4}
\eal
where $\br = r(\cos\phi,\sin\phi)$, $H_m^{(2)}$ denotes the Hankel function of the second kind of order $m$, $\xi=v_\text{s}/\delta_A$ and $\delta_A=\omega_\text{Dirac}-\omega_A$. The prefactors are given by
\bal\label{Prefactors}
\qquad \quad\; P^{\pm}(\br) = \fr{\mathcal Ac^2|E_0|^2\delta_A}{8\omega_Av_\text{s}^2}\left(  e^{i\bp_{\mbf K}\cdot \br} \pm  e^{i\bp_{\mbf K'}\cdot \br} \right).
\eal
Note that this Green's function expression is only valid when the position of the source ($\br_1$) is at the center of the unit cell of the photonic crystal and the Green's function is sampled at a position that is a lattice vector away ($\br_1 = \br_2 + \bR$). 

It is instructive to reexpress the Green's function in a circularly polarized basis using ${\ket{\sigma_\pm}=\mp(\ket{x}\pm i\ket{y})/\sqrt{2}}$, which yields
\bal\label{Greens}
G_{\sigma_+\sigma_+}(\br)&=&-P^+(\br)H_0^{(2)}(r/\xi ), \\
G_{\sigma_-\sigma_-}(\br)&=&-P^+(\br)H_0^{(2)}(r/\xi ),\\
G_{\sigma_+\sigma_-}(\br)&=& e^{i\phi}P^-(\br)H_1^{(2)}(r/\xi),\\
G_{\sigma_-\sigma_+}(\br)&=&-e^{-i\phi}P^-(\br)H_1^{(2)}(r/\xi ).
\eal 
Note the presence of the {\it winding phases} $e^{\pm i \phi}$ in the off-diagonal terms \cite{Peter2015,Karzig2015,Bettles2017}. These phases are of particular importance, as they give rise to chiral interactions and topological behavior when emitters interact through the photonic Dirac cone. The topological properties of a two-dimensional lattice of emitters embedded in a photonic crystal are discussed in more detail in Ref.~\cite{Perczeletal2018}.

\begin{figure}[h!]
\centering
\includegraphics[width=0.475 \textwidth]{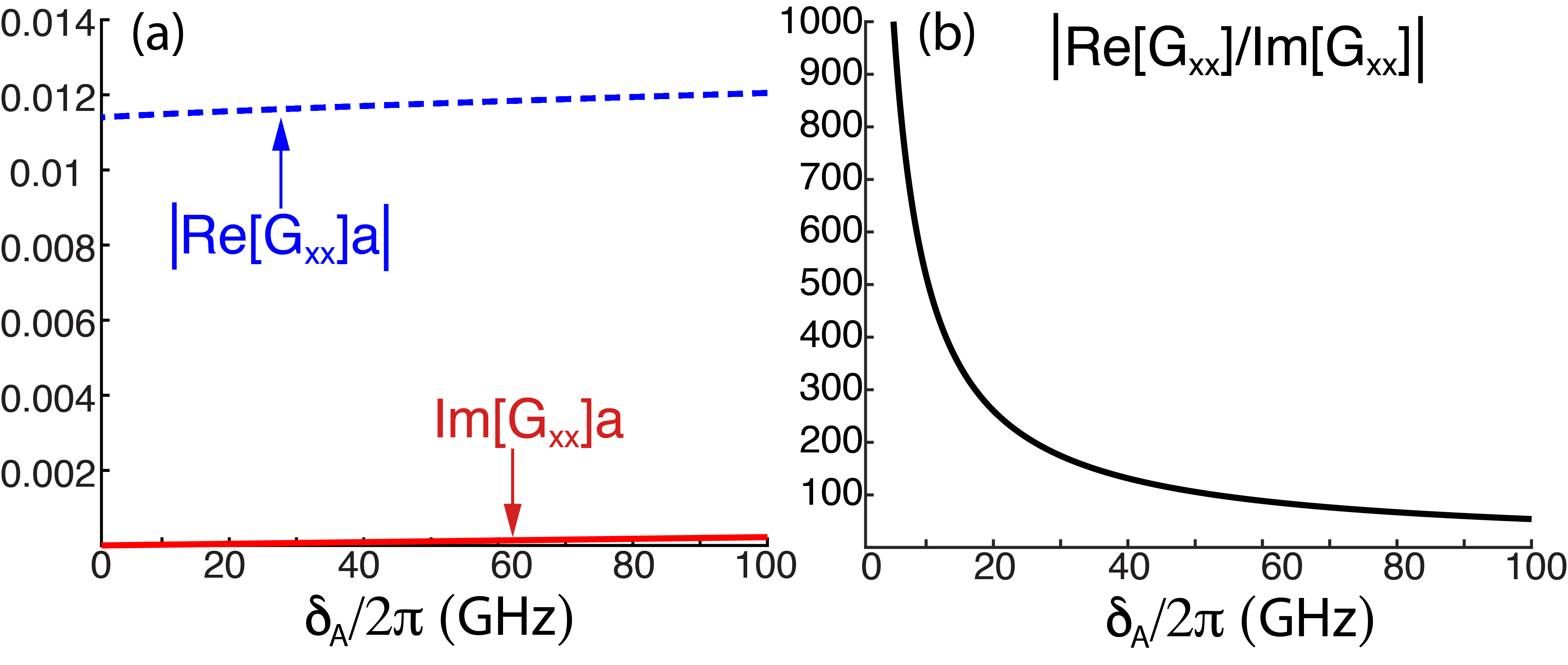}
\caption{
\label{fig:interactions}
(a) Real and imaginary parts of the Green's function as a function of the detuning, $\delta_A$, corresponding to the coherent and dissipative parts of the dipolar interaction, respectively. The dissipative part goes to zero at the Dirac point. (b) Ratio of the coherent and dissipative interactions as a function of the detuning, $\delta_A$. The Green's function is evaluated 5 lattice sites away from the dipole source along the axis aligned with the $\mbf a_+$ lattice vector. The transition wavelength of the atoms is assumed to be $\lambda=738$nm and the Green's function parameters are $|E_0|^2/a^3=0.75$ and $v_\text{s} = 0.3c$.}
\end{figure}

\section{Analysis of the dipolar interaction}\label{InteractionsAnalysis}

In this section we analyze the dipolar interaction mediated by the photonic Dirac cone between atoms. 

Recently, the dipolar interaction mediated by a photonic Dirac cone between atoms was studied in Ref.~\cite{Gonzalez-Tudela2018} using a tight-binding model. It was shown that the emerging atomic interactions are purely coherent, with negligible dissipative terms, and have a long range without exponential attenuation. This stands in sharp contrast with previous results in structured reservoirs, where dissipative terms could only be suppressed at the cost of making the interaction decay exponentially with distance \cite{Shahmoon2013,Douglas2015,Gonzalez-Tudela2015}. As an application of our formalism, we now study the emergence of long-range coherent interactions, when all polarization effects are included, and then provide expressions for the asymptotic behavior of the dipolar interaction as a function of distance.

\reffig{fig:interactions}(a) shows the real and imaginary parts of the Green's function (Eqs.~\eqref{GreensCartesian1}--\eqref{GreensCartesian4}), corresponding to the coherent dipole-dipole interaction (Eq.~\eqref{coop}) and dissipative cooperative decay (Eq.~\eqref{dd}), as a function of the detuning $\delta_A$. The two interacting atoms are assumed to be $x$-polarized and 5 lattice sites away. The dissipative part of the interaction goes to zero as the atomic frequencies become resonant with the Dirac point, whereas the coherent part remains finite. \reffig{fig:interactions}(b) shows the ratio of the coherent and dissipative parts of the interaction as a function of $\delta_A$. Close to the Dirac point, the dipole-dipole interaction can be several orders of magnitude larger than the cooperative decay. Thus the interaction between atoms is almost completely coherent. Note that as $\delta_A \to 0$, the Born-Markov approximation eventually breaks down and the simple description of the dipolar interaction via the dyadic Green's function becomes invalid \cite{Gonzalez-Tudela2018,Perczeletal2018}.


\begin{figure}[h!]
\centering
\includegraphics[width=0.485 \textwidth]{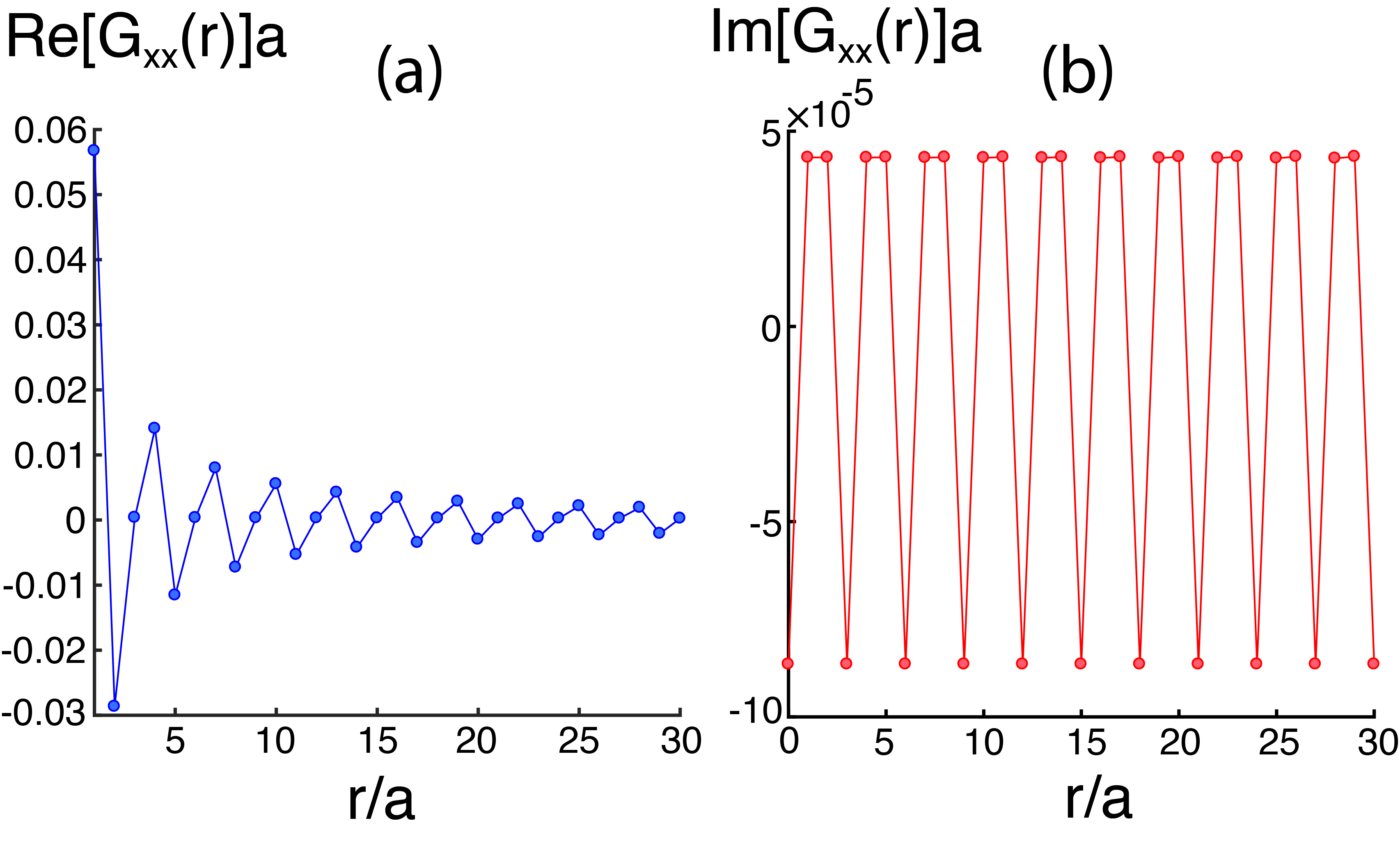}
\caption{
\label{fig:greens_plot}
(a) Dipole-dipole interaction (Re[$G_{xx}(r)$]) and (b) cooperative decay (Im[$G_{xx}(r)$]) as a function of distance between two $x$-polarized atoms. The transition wavelength of the atoms is assumed to be $\lambda=738$nm, the detuning is $\delta_A=19.5$GHz, and the Green's function parameters used are $|E_0|^2/a^3=0.75$ and $v_\text{s} = 0.3c$.} 
\end{figure}

In \reffig{fig:greens_plot} we plot, as a function of distance, the real and imaginary parts of the Green's function, which correspond to the dipole-dipole interaction and cooperative decay between two $x$-polarized atoms, respectively (see Eqs.~\eqref{coop}-\eqref{dd}).
The dipole-dipole interaction initially decreases rapidly away from the source, before transitioning to a slowly decaying oscillatory regime. Note that in Fig.~\ref{fig:greens_plot}(b), the imaginary part of the Green's function is non-zero at the source, which corresponds to the decay rate of the $x$-polarized excited state into the modes of the slab (see \refeq{spontaneousdecay}). 



We can obtain the asymptotic analytic behavior of the interaction, by re-expressing the Hankel functions of the Green's function in terms of Bessel functions as
\bal\label{HankelDef}
H_m^{(2)}(x) = J_m(x) - iY_m(x),
\eal
and observing that for $x \gg 1$ to leading order we get
\bal
J_m(x) \sim \fr{1}{\sqrt{x}}\quad\text{and}\quad Y_m(x) \sim \fr{1}{\sqrt{x}}.
\eal
Thus, both the diagonal and off-diagonal terms of the Green's function fall off as $\sim1/\sqrt{r}$ with distance. For completeness, we also note that when $x\ll 1$, we obtain $J_m(x)\sim 1$, $Y_0(x)\sim \ln(x)$ and $Y_1(x)\sim 1/x$. 

These results confirm that the photonic Dirac cone mediates long-range coherent interactions between emitters without exponential attenuation. Such long-range coherent interactions are of significant importance in quantum information processing, as they can be utilized to entangle \cite{Shahmoon2013,Perczel2018,Gonzalez-Tudela2018} and perform deterministic phase gates between distant atomic qubits \cite{Dzsotjan2010}. Long-range coherent interactions are also advantageous for engineering nearly flat bands in topological quantum optical systems, which have important applications in exploring exotic many-body phases \cite{Perczeletal2018}.


\section{Conclusion}\label{Conclusion}
In conclusion, we have developed an analytic formalism to describe dipole radiation near a realistic two-dimensional photonic crystal slab with a photonic Dirac cone. We derived a closed-form dyadic Green's function and showed that it mediates interactions that feature winding phases, which are key ingredients for engineering topology in photonic systems. As an example, we studied the long-range coherent interactions mediated by the Dirac cone between emitters, which have important applications in quantum information processing. We believe that these results will enable further, rigorous studies of the behavior of atomic emitters interacting via photonic Dirac cones.  

\section{Acknowledgments}

We would like to thank Mihir Bhaskar, Johannes Borregaard, Ruffin Evans, Alejandro Gon\'alez-Tudela, Efraim Shahmoon, Denis Sukachev, Dominik Wild and Peter Zoller for valuable discussions. This  work  was  supported through the National Science Foundation (NSF), the Center for Ultracold Atoms, the Air Force Office of Scientific  Research  via  the  MURI,  the  Vannevar  Bush Faculty Fellowship and DOE. Some of the computations in this paper were performed on the Odyssey cluster supported by  the  FAS  Division  of  Science,  Research  Computing Group at Harvard University. J. P. acknowledges support from the Dr. Elem\'er and \'Eva Kiss Scholarship Fund.

\section*{Appendix}

\appendix\label{appendix}

In this Appendix, we describe the analytic calculations behind obtaining the Green's function of Eqs.~\eqref{GreensCartesian1}-\eqref{Prefactors}.  

Given the approximations described in Section~\ref{GreensFunction}, we can rewrite \refeq{integral1} as 
\bal
G_{\alpha\beta}(\br_2) &\approx& \fr{\mathcal{A}c^2}{8\pi^2\omega_Aa^3} \bigg(I_{\mbf K,\alpha\beta}^{(+)} + I_{\mbf K,\alpha\beta}^{(-)} \nonumber\\
&&\qquad\qquad+ I_{\mbf K',\alpha\beta}^{(+)} + I_{\mbf K',\alpha\beta}^{(-)} \bigg),
\eal
where the four terms in the brackets correspond to the contributions from the upper and lower bands of the Dirac cones at the $\mbf K$ and $\mbf K'$ points. The integrals at the $\mbf K$ point are given by 
\be
I_{\mbf K,\alpha\beta}^{(\pm)}=\int\! d^2\bp\fr{u_{\bp,\alpha}^{(\mbf K,\pm)*}(\br_1)\,u_{\bp,\beta}^{(\mbf K,\pm)}(\br_1)}{\omega_A-\omega_\bp^{(\mbf K, \pm)}}e^{i\bp\cdot \br_{12}},
\ee
and an analogous expressions holds at the $\mbf K'$ point. The evaluation of these integrals follows closely the methods outlined in Ref.~\cite{Gonzalez-Tudela2015}.

First, we substitute \refeq{Dispersion} into the denominator, and after changing changing variables to $\bk = \bp -\bp_\mbf{K}$, we obtain 
\bal\label{Iintegral1}
I_{\mbf K,\alpha\beta}^{(\pm)}=e^{i\bp_\mbf{K}\cdot \br_{12}}\!\!\int\!d^2\bk\fr{u_{\bp_\mbf{K}+\bk,\alpha}^{(\pm)*}\,u_{\bp_\mbf{K}+\bk,\beta}^{(\pm)}}{\omega_A-\omega_\text{Dirac}\mp v_\text{s}|\bk|}e^{i\bk\cdot \br_{12}},\quad
\eal
where for notational convenience we have made the positional argument of $u_{\bp_\mbf{K}+\bk,\alpha}^{(\pm)*}(\br_1)$ implicit. 

We proceed by parameterizing the position vector $\br_{12}$ and momentum vector $\bk$ as follows
\be
\br_{12} = r(\cos\phi,\sin\phi),
\ee
\be
\bk = k(\cos \Phi_\mbf{K},\sin\Phi_\mbf{K}),
\ee
where $\phi$ is measured from the $x$-axis and $\Phi_\mbf{K}$ is given by \refeq{PhiK}. Using these parameterizations, we obtain
\be\label{parameterization}
\bk\cdot\br_{12}=k\,r\cos(\phi-\Phi_\mbf{K}),
\ee
and substituting this expression into \refeq{Iintegral1} yields
\bal\label{masterI}
I_{\mbf K,\alpha\beta}^{(\pm)}\!=e^{i\bp_{\mbf K}\cdot \br_{12}}\!\!\!\int\!\!d^2\bk\fr{u_{\bp_{\mbf K}+\bk,\alpha}^{(\pm)*}\,u_{\bp_\mbf{K}+\bk,\beta}^{(\pm)}e^{ikr\cos(\phi-\Phi_\mbf{K})}}{\omega_A-\omega_\text{Dirac}\mp v_\text{s}|\bk|}.\quad\;\,\,\,\,
\eal
First, we evaluate this expression for the top band when $\alpha\beta=xx$. Substituting \refeq{PolarizationStructureK} and \refeq{normalizedU} into \refeq{masterI}, we obtain
\bal\label{Ixx}
I_{\mbf K,xx}^{(+)}&=&a^3|E_0|^2e^{i\bp_{\mbf K}\cdot \br_{12}}\!\int\limits_0^{k_c}\!kdk\fr{\Lambda(\phi,kr)}{-\delta_{A}- v_\text{s}k},
\eal
where we take the limit of the integral to be $k_c=|\bp_\bK|$ for simplicity. We have also used the notation $\delta_{A}=\omega_\text{Dirac}-\omega_A$ and $|\bk|=k$, and defined the expression
\bal
\Lambda(\phi,kr) = \int\limits_0^{2\pi}d\Phi_\bK\;\sin^2\left(\fr{\Phi_\mbf{K}}{2}-\fr{\pi}{4}\right)e^{ik\,r\cos(\phi-\Phi_\bk)},\qquad
\eal
which, after changing variables to $\phi-\Phi_\mbf{K}=\theta$, evaluates to
\bal
\Lambda(\phi,kr) =  \pi J_0( kr)  - 
   i \pi \sin\phi\;
J_1( kr).
\eal
Next, we substitute this expression into \refeq{Ixx} and change variables to $q = v_\text{s}k/\delta_A $. We also extend the limit of integration with respect to $q$ to infinity based on the observation that $v_\text{s} k_c/\delta_A\gg 1$ when $\delta_A$ is small. Performing the resulting integral yields
\bal
&&I_{\mbf K,xx}^{(+)}= - \fr{a^3|E_0|^2\delta_A\pi}{2v_\text{s}^2}e^{i\bp_\bK\cdot \br_{12}}\Bigg(\fr{2}{r/\xi} +\pi Y_0(r/\xi)\nonumber\\
&& - \pi H_0(r/\xi) +i \pi \sin\phi\big[Y_1(r/\xi)+H_{-1}(r/\xi)\big]\Bigg),\quad
\eal
where $Y_m(r)$ and $H_m(r)$ are Bessel and Struve functions of order $m$, respectively, and we have defined $\xi=v_\text{s}/\delta_A$, which sets the length scale of the interaction mediated by the Green's function.

For the bottom band, we need to evaluate the integral 
\bal\label{IxxBottom}
I_{\mbf K,xx}^{(-)}&=&a^3|E_0|^2e^{i\bp_{\mbf K}\cdot \br_{12}}\!\int\limits_0^{k_c}\!kdk\fr{\Lambda'(\phi,kr)}{-\delta_{A}+ v_\text{s}k},
\eal
where
\bal
\Lambda'(\phi,kr) &=& \int\limits_0^{2\pi}d\Phi_\bK\;\sin^2\left(\fr{\Phi_\mbf{K}}{2}+\fr{\pi}{4}\right)e^{ik\,r\cos(\phi-\Phi_\bk)},\nonumber\\
&=&  \pi J_0( kr)  + i \pi \sin\phi\; J_1( kr).
\eal
This integral is singular, since the denominator can be zero. To make the integral well-defined, we introduce a small imaginary component $+i\vare$, where $\vare>0$, into the denominator to move the pole off the real axis. This corresponds to only considering the outgoing waves that arise from the point source, making the Green's function causal \cite{Chew1995}. Then, after changing variables and extending the limit of integration as for the top band, we use the following identity
\bal
\fr{1}{x\pm i\vare}=\text{P}\left[\fr{1}{x}\right]\mp i\pi\delta(x),
\eal
to obtain
\bal
I_{\mbf K,xx}^{(-)}&=& - \fr{a^3|E_0|^2\delta_A\pi}{2v_\text{s}^2}e^{i\bp_\bK\cdot \br_{12}}\Bigg(-\fr{2}{r/\xi} +\pi Y_0(r/\xi)\nonumber\\
& +& \pi H_0(r/\xi) +i \pi \sin\phi\big[Y_1(r/\xi)-H_{-1}(r/\xi)\big],\nonumber\\
&+&2i\big[\pi J_0( r/\xi)  + i \pi \sin\phi\; J_1( r/\xi)\big]\Bigg).
\eal
Finally, adding up the contributions from the top and bottom bands, we obtain
\bal\label{IKxxsum}
I_{\mbf K,xx}^{(+)} + I_{\mbf K,xx}^{(-)} = \fr{a^3|E_0|^2\delta_A\pi^2}{v_\text{s}^2}e^{i\bp_\bK\cdot \br_{12}}\Bigg(- iH_0^{(2)}(r/\xi)\nonumber\\
+\sin\phi H_1^{(2)}(r/\xi) \Bigg),\qquad\qquad
\eal
where we have used the the definition for the Henkel function of the second kind of order $m$ in terms of Bessel functions (Eq.~\eqref{HankelDef}). 

The integrals near the $\mbf K'$ point can be evaluated using analogous calculations. The only difference is that we need to use $\bp_{\mbf K'}$ instead of $\bp_\mbf{K}$ and the electric field modes are different. The results are identical to those in \refeq{IKxxsum} with the replacements $\bp_{\mbf K}\to \bp_{\mbf K'}$ and $\sin\phi \to -\sin\phi$ yielding
\bal\label{IKPrimexxsum}
I_{\mbf K',xx}^{(+)} + I_{\mbf K',xx}^{(-)} = \fr{a^3|E_0|^2\delta_A\pi^2}{v_\text{s}^2}e^{i\bp_\bK'\cdot \br_{12}}\Bigg(- iH_0^{(2)}(r/\xi)\nonumber\\
-\sin\phi H_1^{(2)}(r/\xi) \Bigg).\qquad\qquad
\eal
Substituting all four terms into \refeq{integral1}, we obtain 
\bal\label{GxxAppendix}
G_{xx}(r,\phi)&=&  \sin\phi\, P^{-}(\br)  H_1^{(2)}( r/\xi)\nonumber\\
&&-iP^{+}(\br) H_0^{(2)}(r/\xi),\quad\;\,\,\qquad\;\;
\eal
where
\bal
\qquad \quad\; P^{\pm}(\br) = \fr{\mathcal Ac^2|E_0|^2\delta_a}{8\omega_Av_\text{s}^2}\left(  e^{i\bp_{\mbf K}\cdot \br} \pm  e^{i\bp_{\mbf K'}\cdot \br} \right).\qquad
\eal
The Green's function for the cases when $\alpha\beta = yy$, $\alpha\beta = xy$ and $\alpha\beta = yx$ can be calculated analogously. In particular, the final result for $G_{yy}$ can be obtained by making the substitution $\sin\phi \to -\sin\phi$ in \refeq{GxxAppendix} to get
\bal\label{GyyAppendix}
G_{yy}(r,\phi)&=&  -\sin\phi\, P^{-}(\br)  H_1^{(2)}( r/\xi)\nonumber\\
&&-iP^{+}(\br) H_0^{(2)}(r/\xi).\quad\;\,\,\qquad\;\;
\eal
The expression for $G_{xy}$ is given by
\bal
G_{xy} (r,\phi)&=&  \cos\phi \, P^{-}(\br)H_1^{(2)}( r/\xi),\quad\;
\eal
and we note that
\bal
G_{yx} (r,\phi)&=& G_{xy} (r,\phi).
\eal
These results are summarized in Eqs.~\eqref{GreensCartesian1}-\eqref{Prefactors}.

\end{document}